\documentclass[11pt]{article}
\usepackage{mathtext}
\usepackage[dvips]{graphicx}

\begin{document}
\title{ON CIRCULAR ORBITS \\ IN EINSTEIN'S THEORY OF GRAVITATION\footnote{Published in soviet physical journal JETP (1949, Vol. 19, No 10, P. 951-952) as Letters to Editorial Board.}}

\author{S.A. Kaplan}
\date{ }


\maketitle

\vspace{-6mm}

\centerline{Astronomical Observatory in Lviv\footnote{Complete title of the institution ``Astronomical observatory of Ivan Franko State (now National) University of Lviv}}


\vspace{6mm}

In papers by Einstein \cite{E} and Oppenheimer \cite{O} it was shown that for some types of physical configurations in Einstein's gravitational field there is the upper finite limit of the density of matter, and the lower limit of radii for these configurations  which differs from Schwarzschild's singularity. For radii less than this limit, the configurations became unstable. 

In this note we will show that the set of the circular orbits in Einstein's gravitational field (more exactly, in spherically symmetric static Schwarzschild's field) has the same property: namely, the circular orbits with a radius less than certain limit are gravitationally unstable.

It is known that the interval for the spherically symmetric static Schwarz\-schild field can be written as 

\begin{equation}
ds^2=\gamma dt^2 -\gamma^{-1} dr^2 - r^2 (\sin^2\theta d\varphi^2 + d\theta^2), \quad \gamma=1-\frac{2m}{r^2}.
\end{equation}
Here we use the gravitational relativistic units, i. e. the gravitational constant
$G =1$ and the velocity of light $c=1$; $m$ is the mass of the central body.

For determination of the trajectory we use the Hamilton-Jacobi equation
\begin{equation}
g^{ik}\frac{\partial{S}}{\partial{x^i}} \frac{\partial{S}}{\partial{x^k}}=m_1^2.
\end{equation}
Here $S$ is action and $m_1$ is the mass of the body moving in the orbit
($m_1\ll m$). For the circular orbit $S$ does not depend on the radius vector $r$. Therefore, 
\begin{equation}
S= - Et + M\varphi,
\end{equation}
where $M$ is the angular momentum. Substituting (3) and (1) into (2) we find
\begin{equation}
E^2 = \gamma \left\{m_1^2 + \frac{M^2}{4m^2} (1-\gamma)^2\right\}.
\end{equation}
Differentiating (4) by $\gamma$ and equating the derivative to zero  (that is, finding the minimum of energy for given $M$) we obtain the expression for $M$ depending on radius of the orbit
\begin{equation}
M^2 = \frac{4m^2 m_1^2}{(3\gamma-1)(1-\gamma)}.
\end{equation}
It follows from this formula that for $1\ge \gamma\ge 1/3$, i. e. the angular momentum is equal to infinity as for the infinitely distant orbit and for the orbit with $r=3m$. We recall that the radius of Schwarz\-schild's singularity is
equal to $2m$. It is necessary to remember that $r$ is not an invarian value. In Schwarz\-schild's field the invariant value is the circuit $l=2\pi r$, but for convenience in the following we will use the "radius" $r$.

Thus, in the interval between the spheres with the radius $3m$ and Schwa\-rzschild's singularity there is not a single real circular orbit. (The interval for the corresponding orbits in this region is spatial).

Substituting (5) into (4) we obtain
\begin{equation}
E^2 = m_1^2 \frac{2\gamma^2}{3\gamma-1}.
\end{equation}
This formula gives the energy of a body in the circular orbit depending on the distant from the center. Hence it follows that, firstly, in the orbit with $r=3m$ the energy $E$ is equal to infinity, i. e. the body in this orbit should move with the speed of light (this result was obtained by Einstein); second, $E=m_1$ both for $\gamma=1$, and for $\gamma=1/4$, i. e. at $r\to \infty$ and $r=4m$. Hence, the energy of the body in the orbit with $r=4m$ is equal to the energy in the infinitely distant orbit. Thus, in the interval between spheres with $r=3m$ and $r=4m$ the energy for the circular orbits is grater then the energy in infinity, and therefore these orbits are unstable.

Differentiating (6) we find that the value of $E$ is minimum at $\gamma=2/3$, i. e. at $r=6m$. For this value $r$
\begin{equation}
E_{min} = m_1^2 \sqrt{8/9}, \quad M_{min} = m_1 m \sqrt{12}.
\end{equation}
Hence it follows that for $1 \ge E/m_1 > \sqrt{8/9}$ there are two circular orbits with the same value of $E$ but with different values of $M$. One of them is outside the sphere with $r=6m$. This orbit is stable and can be named the Newtonian orbit. The second orbit is inside the sphere with $r=6m$ and is unctable.

The orbit with $r=6m$ is the minimal stable circular orbit in Schwarz\-schild's
gravitatiobal field. It is known that in the Newtonian gravitational field there is no minimum stable circular orbit (more exactly, all circular orbits are stable up to the center, where their energy becomes negative infinity).

\end{document}